# Magnetosomes in Nature, Biomedicine and Physics

N.A. Usov

*Pushkov Institute of Terrestrial Magnetism, Ionosphere and Radio Wave Propagation, Russian Academy of Sciences, IZMIRAN, 108480, Troitsk, Moscow, Russia*

*E-mail: usov@obninsk.ru*

**Abstract** Magnetotactic bacteria synthesize linear chains of magnetite nanoparticles within their bodies, which allow the bacteria to navigate the Earth's magnetic field in search of the best habitat. Biogenic magnetite particles, called magnetosomes, are very promising for use in biomedicine. Magnetosome chains have also been found in ancient fossils and sediments. The study of magnetofossils provides valuable information about the Earth's biological past. The presence of biogenic magnetite in ancient rock samples can be detected by measuring ferromagnetic resonance spectra, first-order magnetization reversal curves, or quasi-static hysteresis loops. Theoretical analyses of these experiments generally assume that magnetosomes are spherical nanoparticles, although the shape of some types of magnetosomes is close to spheroidal one. In this work, simple formulas for describing the magneto-dipole interaction of oriented spheroids are obtained and quasi-static hysteresis loops of randomly oriented magnetosome chain assembly consisting of elongated spheroids are calculated.

## INTRODUCTION

Magnetotactic bacteria are unique microorganisms that synthesize inside their bodies peculiar skeletons in the form of linear chains of magnetite nanoparticles [1-5]. Magnetic nanoparticles that form a linear chain are called magnetosomes. From the biological point of view, it is interesting to understand for what purposes the bacterium builds this magnetic skeleton, as well as how the synthesis of magnetic nanoparticles inside a living organism occurs [1,2,5-7]. From the physical point of view, it is important that the production of magnetic nanoparticles occurs under physicochemical conditions strictly controlled by the bacterial organism. As a result, the magnetite nanoparticles synthesized by the bacterium have a perfect crystalline structure, a narrow distribution of particle sizes and shapes, and a high saturation magnetization, $M_s = 85 – 90$ emu/g, close to that of a magnetite single crystal [8]. The number of magnetosomes in a linear chain can be different, and varies usually from 10 to 30 nanoparticles [1-4].

The sizes of magnetosomes are almost always smaller than the single-domain diameter, which for a spherical magnetite nanoparticle is $D_c = 70.4$ nm [9]. Consequently, magnetosomes with characteristic diameters $D = 20 – 50$ nm are small permanent magnets. Neighboring particles spontaneously magnetize each other since they are spatially arranged in the form of a linear chain at a close distance from each other. As a result, despite the relatively small value of the cubic magnetic anisotropy constant of magnetite, $K_c = -10^5$ erg/cm$^3$ [10], due to the strong magneto-dipole interaction in the chain, the linear magnetic skeleton of the bacterium has a virtually permanent magnetic moment, which is weakly affected by thermal fluctuations at room temperature.

An important geometrical parameter of the chain is an average distance $a_c$ between the centers of particles in the chain, since this distance determines the amplitude of the dipolar field $H_{dip}$, acting on a particle in the chain. Based on the transmission electron microscope (TEM) images [1,2,11,12], it can be concluded that the nearest distance between the surfaces of neighboring spherical particles is the sum of the thicknesses of the magnetosome shells $2T_{en}$. Here $T_{en} = 4 – 6$ nm is the characteristic thickness of the lipid magnetosome shell that weakly depends on the nanoparticle diameter. Thus, the average distance between the particle centers in a chain is $a_c = D + 2T_{en}$. As a result, the characteristic value of the dipolar magnetic field $H_{dip}$, which acts on an individual particle far from the chain ends, is estimated as $H_{dip} = (2\pi/3) M_s \zeta(3)/(1 + 2T_{en}/D)^3$ [13], where $\zeta(3) \approx 1.2$ is the value of Riemann zeta function [14]. It is assumed [1-4] that the presence of a non-magnetic shell protects magnetite nanoparticles from oxidation and transformation into maghemite.

Another interesting question is how individual magnetosomes are spatially oriented with respect to the linear chain axis. Magnetite nanoparticles have a cubic type of magnetic anisotropy, and since the magnetic anisotropy constant of magnetite is negative, a quasi-spherical magnetite particle has 8 easy magnetization axes [10]. It seems reasonable to assume that the orientation of the anisotropy axes of an individual particle with respect to the chain axis can be random. However, a number of studies show [15,16] that usually one of the cubic easy anisotropy axes of each nanoparticle is directed along the linear chain. It is clear that in such a case the stability of the total magnetic moment of the bacterium in the direction of the chain axis increases. A number of studies also emphasize [1,2,7] that in order to increase the mechanical stability of the linear arrangement of particles, the linear chain as a whole is dressed in a strong framework of surrounding non-magnetic tissues.

Due to the above factors, an adult bacterium, which has a developed magnetic skeleton of 10-20 magnetosomes in a chain, should easily feel the influence of the weak magnetic field of the Earth, which is about 0.5 Oe [17], on its orientation in space. Magnetotactic bacteria live in an aquatic environment, in lakes and rivers, as well as at great depths in oceans, and the correct orientation in space helps the bacterium to find the best living conditions [1-4]. It is amazing how effectively biological selection works in nature, allowing organism to modify its structure in a best way from a physical point of view.

It is the high quality and homogeneity of the geometric characteristics of magnetic nanoparticles, which are still very difficult to ensure when synthesizing magnetic nanoparticles in laboratory, that promotes the use of magnetosomes in biomedicine, in particular, in magnetic hyperthermia [13,18-24]. Another important factor is the low toxicity of magnetite for the human body, since magnetite is a part of red blood cells [18]. A very high value of specific absorption rate, SAR = 960 W/g, was first achieved by R. Hergt with co-workers [19] for an oriented assembly of magnetosome chains in an alternating magnetic field with a frequency 410 kHz and an amplitude of 10 kA/m. Later, even higher SAR values were obtained on oriented assemblies of various types of magnetosome chains [23,24]. Numerical simulation shows [13] that in order to obtain high SAR values for magnetosome chain assembly in alternating magnetic field of moderate amplitude, it is necessary to provide an optimal ratio $a_c/D$ of the distance $a_c$ between the particle centers to the particle diameter $D$. The prospects for using magnetosomes in biomedicine have been demonstrated, in particular, by E. Alphandéry with co-workers [20,22].

It is well known [1-4] that various types of magnetotactic bacteria are widespread in nature and are found in many lakes and rivers on land. At the same time, these bacteria are rather capricious and reproduce well only in clean natural water bodies. In particular, special conditions are required to grow such bacteria in a laboratory. It is important that the presence of magnetotactic bacteria residues in the bottom silt allows researchers to judge the evolution of the pollution content in a water body over long periods of time [25,26]. The uniqueness of magnetotactic bacteria is that their former presence in the water body can be detected by the occurrence of linear chains of magnetic particles in water sediments. At the same time, the biological material of ordinary bacteria completely disappears in the natural environment in a short time.

Interestingly, the remains of ancient magnetotactic bacteria have also been found in natural sediments on the bottom of seas and oceans, the age of which is estimated in some cases to be tens of millions of years [25-27]. Magnetotactic bacteria are possibly among the most ancient organisms producing nanosized magnetite intracellularly. Identifying bacterial magnetofossils in ancient sediments is important for assessing life evolution over geological times. Thus, the study of magnetofossils in bottom sediments is of interest to microbiologists, ecologists, and researchers of other specialties. However, there is a problem of detecting microscopic bacterial remains in the natural sediments, which usually occupy a large volume and can extend for tens of centimeters and even meters [25-27]. The most informative technique for detecting magnetofossils, i.e. transmission electron microscopy (TEM), is obviously expensive. In addition, when preparing samples for TEM analysis, destruction or violation of the integrity of the magnetotactic bacteria remains under study is possible [11]. Therefore, alternative techniques for detecting magnetotactic bacteria remains have been actively developed recently, such as measuring the ferromagnetic resonance spectra (FMR) [28-37] of bottom sediments, analyzing their first-order magnetization reversal curves [38-41] and studying quasi-static hysteresis loops of various assemblies of magnetosome chains [42-46].

Note that in theoretical calculations of FMR spectra [34,37], or quasi-static hysteresis loops [42,46] of magnetosome chain assemblies it is usually assumed that magnetosomes are quasi-spherical particles, since this allows one to describe the magneto-dipole (MD) interaction of single-domain nanoparticles by a simple formula. At the same time, magnetosomes synthesized by some types of magnetotactic bacteria have a shape close to an elongated spheroid [1,2,47,48]. For single domain particles of ellipsoidal or spheroidal shape there are only complicated expressions represented by multidimensional integrals over volumes or over particle surfaces [49]. The MD interaction in oriented assembly of single-domain spheroids was recently studied in Ref. 50. In this paper we apply this approach to calculate quasi-static hysteresis loops of randomly oriented assembly of magnetosome chains consisting of elongated spheroids.

## MD INTERACTION OF ELONGATED SPHEROIDS

Let us consider a linear chain of elongated single-domain spheroids with an aspect ratio $a/b > 1$, where $a$ and $b$ are the longitudinal and transverse semi axes of the spheroid, respectively. Without loss of generality, one can assume that the linear chain of spheroids is oriented along the $Z$ axis of the Cartesian coordinates. The energy of the magneto-dipole interaction of two spheroids can be written as a six-dimensional integral over the particle volumes, $V = 4\pi ab^2/3$, [49,50]

$$W_m = M_s^2 \int_V (\vec{\alpha}_1 \vec{\nabla}_1) \int_V (\vec{\alpha}_2 \vec{\nabla}_2) \frac{dv_1 dv_2}{|\vec{r}_1 - \vec{r}_2|} = \sum_{i,j} B_{ij} \alpha_i^{(1)} \alpha_j^{(2)}. \qquad (i,j = x, y, z), \qquad (1)$$

Here $B_{ij}$ are the elements of the MD interaction matrix for spheroidal nanoparticles, $\vec{\alpha}_1$ and $\vec{\alpha}_2$ are the unit magnetization vectors of the first and second single-domain nanoparticle, respectively.

Let the centers of two spheroids be located at the points $(0,0,0)$ and $(0,0,Z_0)$. It is convenient to introduce the reduced distance $Z_0/a$ between the particle centers. Based on the formulas of Ref. 37, it can be shown that in Eq. (1) only the diagonal elements of the matrix $B_{ij}$ are nonzero, that is, $B_{xx}$, $B_{yy}$ and $B_{zz}$. Furthermore, $B_{xx} = B_{yy}$ due to the symmetry of the problem. For the matrix element $B_{xx}$, the following representation can be obtained [50]

$$B_{xx}(Z_0/a) = 2M_s^2 a^2 b \int dS_1 \int dS_2 \left\{ \frac{1}{\sqrt{(t_1 - t_2)^2 + q_{xx}^2}} - \frac{1}{\sqrt{(t_1 + t_2)^2 + q_{xx}^2}} \right\}, \qquad (2a)$$

where

$$q_{xx}^2 = (\rho_1 \cos\varphi_1 - \rho_2 \cos\varphi_2)^2 + (a/b)^2 (\rho_1 \sin\varphi_1 - \rho_2 \sin\varphi_2 - Z_0/a)^2. \qquad (2b)$$

Similarly, for the matrix element $B_{zz}$ one has

$$B_{zz}(Z_0/a) = M_s^2 b^3 \int dS_1 \int dS_2 \left\{ \frac{1}{\sqrt{(a/b)^2 (t_1 - t_2 - Z_0/a)^2 + q_{zz}^2}} - \frac{1}{\sqrt{(a/b)^2 (t_1 + t_2 + Z_0/a)^2 + q_{zz}^2}} - \right.$$

$$\left. \frac{1}{\sqrt{(a/b)^2 (t_1 + t_2 - Z_0/a)^2 + q_{zz}^2}} + \frac{1}{\sqrt{(a/b)^2 (t_1 - t_2 + Z_0/a)^2 + q_{zz}^2}} \right\}, \qquad (3a)$$

where

$$q_{zz}^2 = \rho_1^2 + \rho_2^2 - 2\rho_1 \rho_2 \cos(\varphi_1 - \varphi_2). \qquad (3b)$$

In Eqs. (2), (3) the variables $t_1 = \sqrt{1-\rho_1^2}$ and $t_2 = \sqrt{1-\rho_2^2}$, and the integration is performed over the area of circles of unit radius, so that $dS_1 = \rho_1 d\rho_1 d\varphi_1$, $dS_2 = \rho_2 d\rho_2 d\varphi_2$. Since the volumes of spheroidal nanoparticles do not intersect by condition, the integrals (2), (3) have no singularities in the denominators of the integrands and can be calculated using standard numerical integration programs. In this paper the matrix elements $B_{xx}$ and $B_{zz}$ are calculated numerically for fixed values of the spheroid aspect ratio $a/b$ = 1.25; 1.5; 2.0; 3.0 as a function of the dimensionless variable $Z_0/a$ in the interval $2 < Z_0/a \leq 6.0$.

It is also useful to expand Eqs. (2), (3) in a series up to the second order in the parameter $\delta^2 = (a^2 - b^2)/Z_0^2$ [50]. These expressions are given by

$$B_{xx}^{(2)}(Z_0) = \frac{M_s^2 V^2}{Z_0^3}\left(1 + 2.4\frac{a^2-b^2}{Z_0^2} + \frac{216}{35}\left[\frac{a^2-b^2}{Z_0^2}\right]^2 + ...\right); \quad B_{zz}^{(2)}(Z_0) = -2B_{xx}^{(2)}(Z_0). \quad (4)$$

It is easy to see that in the limit $a = b$ all corrections proportional to the parameter $\delta^2$ vanish, so that one obtains zero-th order spherical approximation for the matrix elements, $B_{xx}^{(0)}$ and $B_{zz}^{(0)}$.

In Fig. 1 the numerical values for the matrix elements $B_{xx}$ and $B_{zz}$, obtained using Eqs. (2), (3), are compared with the analytical expressions (4). For completeness, in Fig. 1 the spherical approximations $B_{xx}^{(0)}$ and $B_{zz}^{(0)}$ for these matrix elements are also shown. The values of the matrix elements are normalized to $M_s^2 a^2 b$.

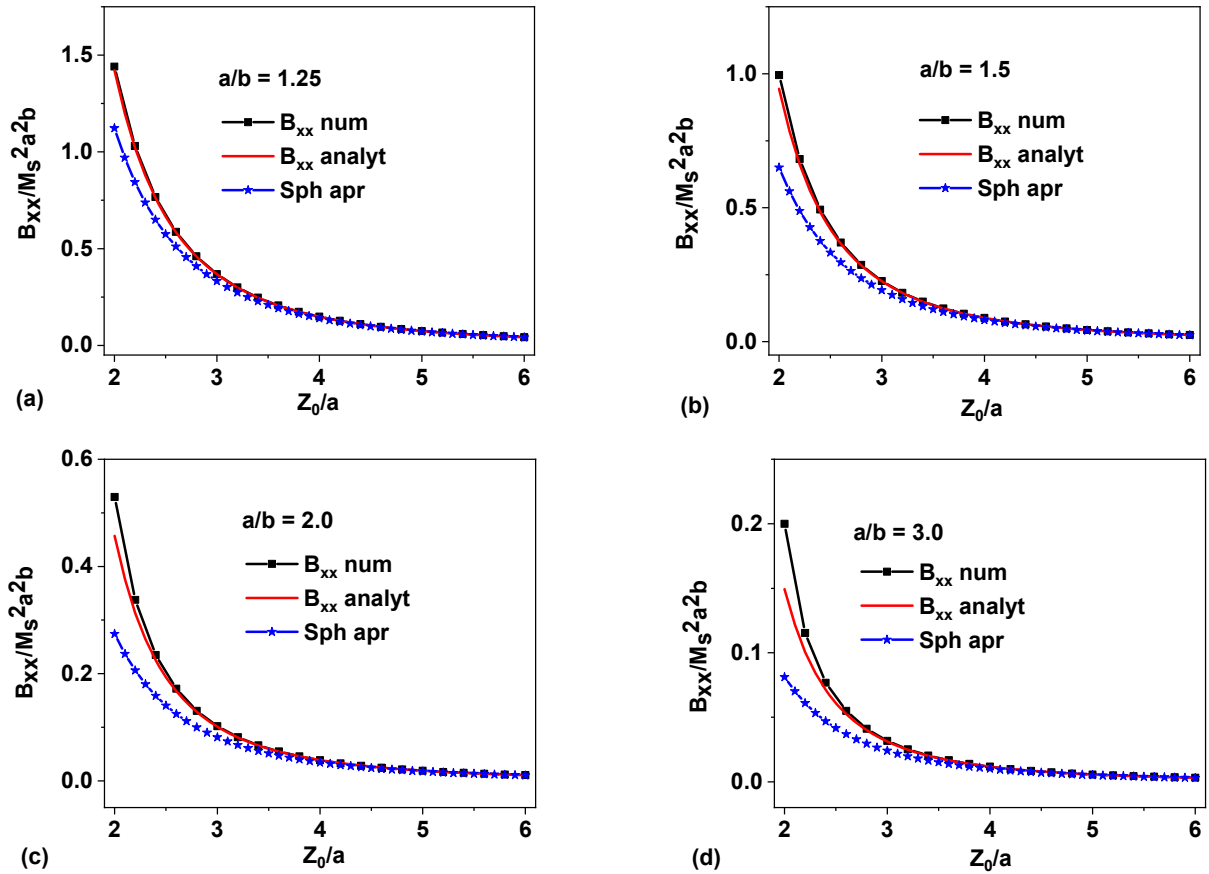

Fig. 1. Comparison of the normalized numerical values for the matrix elements $B_{xx}$ with analytical expressions (4) for various spheroid aspect ratios: a) $a/b$ = 1.25; b) $a/b$ = 1.5; c) $a/b$ = 2.0 and d) $a/b$ = 3.0.

As Fig. 1 shows, the spherical approximation for the elements of the MD interaction matrix of spheroids is rough already for $a/b \geq 1.5$, while the second-order approximation in the parameter $\delta^2$ describes quite accurately the dependence of the matrix elements $B_{xx}$ on the parameter $Z_0/a$, at least for spheroids with aspect ratio $a/b \leq 2.0$. And only for spheroids with aspect ratio $a/b = 3.0$ does some difference arise between the analytical (4) and numerical values (2), (3) of the elements $B_{xx}$ in a small range of values of the variable $Z_0/a \approx 2$. Similar behavior was obtained also for matrix elements $B_{zz}$ as a function of the parameter $Z_0/a$. Analysis of TEM images shows that the shape of many types of elongated magnetosomes can be approximated by spheroids with aspect ratio $a/b \leq 2.0$. Thus, approximation (4) is sufficient for calculating quasi-static hysteresis loops of linear chains of elongated magnetosomes with a moderate aspect ratio.

## QUASISTATIC HYSTERESIS LOOPS

Quasi-static hysteresis loops of linear chains of elongated spheroidal magnetite nanoparticles are calculated in this work by solving the dynamic Landau-Lifshitz-Gilbert equation with phenomenological damping [51,52]. The particle saturation magnetization is taken to be $M_s = 450$ emu/cm$^3$, the cubic magnetic anisotropy constant $K_c = -10^5$ erg/cm$^3$. The magnetic damping parameter is given by $\kappa = 0.5$, the number of particles in the chain is fixed to $N_p = 20$.

The magneto-crystalline anisotropy energy of a linear chain is

$$W_a = K_c V \sum_{i=1}^{N_p} \left( (\vec{\alpha}_i \vec{e}_{1i})^2 (\vec{\alpha}_i \vec{e}_{2i})^2 + (\vec{\alpha}_i \vec{e}_{1i})^2 (\vec{\alpha}_i \vec{e}_{3i})^2 + (\vec{\alpha}_i \vec{e}_{2i})^2 (\vec{\alpha}_i \vec{e}_{3i})^2 \right), \qquad (5)$$

where $(\vec{e}_{1i}, \vec{e}_{2i}, \vec{e}_{3i})$ is a set of orthogonal unit vectors defining the spatial orientation of the cubic easy anisotropy axes of the $i$-th nanoparticle of the chain. In accordance with Refs. 15, 16 it is supposed that the orthogonal sets $(\vec{e}_{1i}, \vec{e}_{2i}, \vec{e}_{3i})$ of various particles are oriented so that one of the easy axes of cubic anisotropy for every particle is parallel to the linear chain axis. Note that other easy anisotropy axes of the particles are randomly oriented in space.

In addition to the magneto-crystalline anisotropy energy (5) for elongated spheroids it is necessary to take into account a shape anisotropy energy contribution

$$W_{sh} = K_{sh} V \sum_{i=1}^{N_p} \left( \alpha_{i,x}^2 + \alpha_{i,y}^2 \right), \qquad (6)$$

where $K_{sh}$ is the shape anisotropy constant. The latter can be calculated as follows [53]

$$K_{sh} = M_s^2 (\pi - 3N_a/4); \qquad N_a = 2\pi \frac{1-\varepsilon^2}{\varepsilon^3} \left( \ln \frac{1+\varepsilon}{1-\varepsilon} - 2\varepsilon \right); \qquad \varepsilon = \sqrt{1-(b/a)^2},$$

where $N_a$ is the demagnetizing factor along the long nanoparticle axis.

The Zeeman energy $W_Z$ of the linear chain in the external uniform magnetic field $\vec{H}$ is given by

$$W_Z = -M_s V \sum_{i=1}^{N_p} (\vec{\alpha}_i \vec{H}). \qquad (7)$$

The direction of the external magnetic field $\vec{H}$ can be specified by the spherical angles $\theta$ and $\varphi$. The magneto-dipole interaction energy $W_m$ of a pair of single-domain spheroidal nanoparticles with aspect ratio $a/b \leq 2.0$ is described by Eqs. (1), (4).

The effective magnetic field $\vec{H}_{ef,i}$ acting on the $i$-th nanoparticle of the chain can be calculated as the derivative of the total chain energy $W$,

$$W = W_a + W_{sh} + W_Z + W_m, \qquad \vec{H}_{ef,i} = -\frac{\partial W}{M_s V \partial \vec{\alpha}_i}. \qquad (8)$$

In accordance with experimental data [11,12], it is assumed that each magnetite nanoparticle is surrounded by a non-magnetic lipid shell with a thickness of $T_{en} = 4$ nm. Therefore, the distance between the centers of particles in a periodic linear chain, which determines the intensity of the magneto-dipole interaction in the chain, is equal to $a_c = 2a + 2T_{en}$.

Due to the partially chaotic orientations of the cubic easy anisotropy axes of the chain particles, to obtain statistically significant results the hysteresis loops of oriented assemblies are averaged over a sufficiently large number $N_{exp} = 40 - 60$ independent realizations of the chains. In each numerical experiment a new linear chain of spheroids with fixed parameters $a$, $b$, $T_{en}$, and $N_p$ is created, the only one of the cubic easy anisotropy axes of every particle being oriented along the chain axis. As a result of such averaging, the dependence of the oriented assembly hysteresis loops on the azimuthal angle $\varphi$ disappears. Therefore, without loss of generality it can be assumed that the direction of the external magnetic field lies in the *XZ* plane and is determined by the polar angle $\theta$. To calculate the hysteresis loops of a randomly oriented assembly of linear chains additional averaging of the hysteresis loops of the oriented assembly is performed over the polar angle $\theta$.

The calculation of the quasi-static hysteresis loop of a chain begins in an external magnetic field, $H = 2000$ Oe, large enough for complete magnetic saturation of the chain. The evolution of the initial magnetization vectors of the particles in a given external magnetic field is traced to the final state, which is assumed to be stable under the condition

$$\max_{1<i<N_p} \left| \vec{\alpha}_i \times \vec{H}_{ef,i} \right| / \left\| \vec{H}_{ef,i} \right\| < 10^{-6}, \qquad (9)$$

where $\vec{\alpha}_i$ and $\vec{H}_{ef,i}$ are the unit magnetization vector and the effective magnetic field of the *i*-th particle of the chain, respectively. After reaching a steady state, the amplitude of the external magnetic field decreases by a small amount $\Delta H = 1$–$2$ Oe and the new steady state is calculated in a similar way. At each step in the magnetic field, the sufficient number of iterations with a small time step is performed until the equilibrium condition of the system magnetization (9) is satisfied.

## RESULTS AND DISCUSSION

Fig. 2a shows the dependence of the quasi-static hysteresis loops of a dilute oriented assembly of linear chains of spheroidal magnetite nanoparticles on the direction of the external magnetic field relative to the common axis of the chains. The magnetosomes have semi axis dimensions $a = 30$ nm, $b = 20$ nm, and an aspect ratio $a/b = 1.5$. The number of particles in the chains is $N_p = 20$, the thickness of the nonmagnetic shell on the particle surface being $T_{en} = 4$ nm. The assembly consists of 40 chains of spheroidal magnetosomes, in which one of the easy axes of cubic magnetic anisotropy is directed along the chain axis whereas the remaining easy axes are randomly oriented.

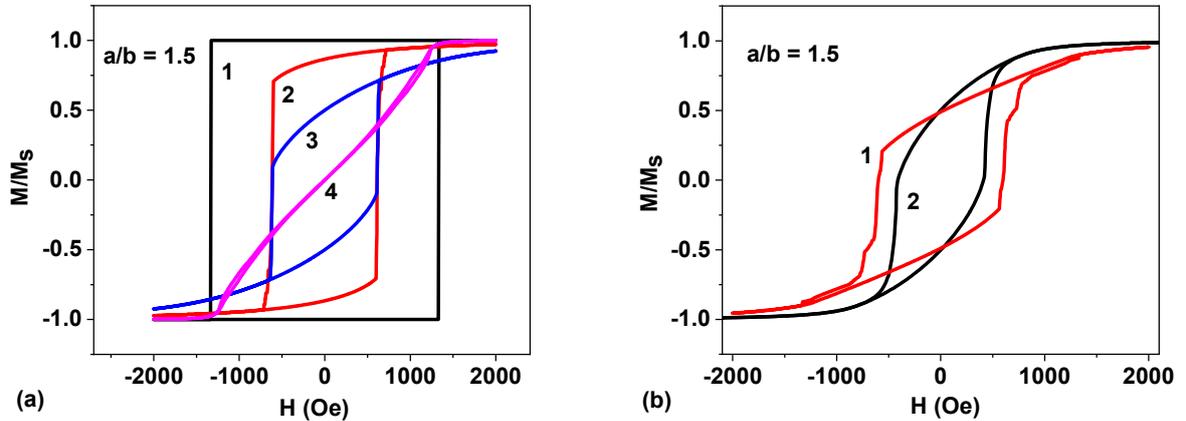

Fig. 2. a) Quasi-static hysteresis loops of oriented assembly of linear chains of elongated magnetosomes with aspect ratio $a/b = 1.5$ for various directions of applied magnetic field with respect to chain axis: 1) $\theta = 0$; 2) $\theta = 30°$; 3) $\theta = 60°$; 4) $\theta = 90°$. b) Comparison of randomly oriented assembly of linear chains of elongated magnetosomes 1) with randomly oriented assembly of non interacting magnetite nanoparticles 2) with the same aspect ratio $a/b = 1.5$.

As Fig. 2a shows, the oriented assembly of chains exhibits generally uniaxial anisotropy, since for a magnetic field direction parallel to the chain axis, $\theta = 0$, the loop is rectangular, with a large coercive force $H_c = 1335$ Oe. On the other hand, for magnetic field directed perpendicular to the chains, $\theta = 90°$, it is close to a linear loop with $H_c = 0$. Number 1) in Fig. 2b marks the hysteresis loop of a dilute non-oriented assembly of linear chains of spheroidal particles with the same magnetic and geometric parameters, having a reduced remanent magnetization $M_r/M_s = 0.5$, and a coercive force $H_c = 598$ Oe. For comparison, number 2) in Fig. 2b marks the hysteresis loop of a randomly oriented assembly of non-interacting spheroidal magnetite nanoparticles with the same aspect ratio $a/b = 1.5$, which has the same remanent magnetization but a lower coercive force, $H_c = 412$ Oe. One notes a considerable difference in the shapes of the both hysteresis loops.

Note that the approach developed in this work can also be used in the theoretical analysis of ferromagnetic resonance spectra and first-order magnetization reversal curves of assemblies of linear chains of spheroidal magnetosomes.